\documentclass[12pt]{iopart}

\usepackage{iopams}
\usepackage{graphicx}
\usepackage{cite}
\begin{document}

\title[]{Phonon-assisted Kondo Effect in a Single-Molecule Transistor out of Equilibrium}

\author{Zuo-Zi Chen$^1$, Haizhou Lu$^1$, Rong L\"{u}$^1$ and Bang-fen Zhu$^{1,2}$}

\address{$^1$~Center for Advanced Study, Tsinghua University,
Beijing 100084, P. R. China $^2$~Department of Physics, Tsinghua
University, Beijing 100084, P. R. China}
\eads{\mailto{zzchen@castu.tsinghua.edu.cn},
\mailto{bfzhu@castu.tsinghua.edu.cn}}
\begin{abstract}
The joint effect of the electron-phonon interaction and Kondo
effect on the nonequilibrium transport through the single molecule
transistor is investigated by using the improved canonical
transformation scheme and extended equation of motion approach.
Two types of Kondo phonon-satellites with different asymmetric
shapes are fully confirmed in the spectral function, and are
related to the electron spin singlet or hole spin singlet,
respectively. Moreover, when a moderate Zeeman splitting is caused
by a local magnetic field, the Kondo satellites in the spin
resolved spectral function are found disappeared on one side of
the main peak, which is opposite for different spin component. All
these peculiar signatures that manifest themselves in the
nonlinear differential conductance, are explained with a clear
physics picture.
\end{abstract}

\maketitle

\section{Introduction}

The Kondo effect as a manifestation of strongly correlations
between conduction electrons and spin impurity has been
extensively studied in the context of quantum dot physics in past
years.\cite{TheoreticalQD1,TheoreticalQD2,KondoBook,ExperimentsQD1,ExperimentsQD2,ExperimentsQD3,ExperimentsQD4}
Recently, its realization in the single-molecule transistor (SMT)
has attracted a lot of
attentions.\cite{Nygard00,ExperimentsSMT1,ExperimentsSMT2,ExperimentsSMT3,ExperimentsSMT4,
FerromaneticLeads1,FerromaneticLeads2} In particular, the latest
experiments in SMT, in which the phonon satellite has been clearly
observed even in the Kondo
regime\cite{KondoSattelites1,KondoSattelites2,KondoSattelites3}
owing to the enhanced electron-vibration coupling in the
molecule,\cite{PhononSMT1,PhononSMT2,PhononSMT3} has stimulated
great interests to investigate the interplay of the
electron-phonon interaction (EPI) and the Kondo effect on the
transport through the SMT.

According to our recent investigation on the spectral function of
the SMT in the absence of the Coulomb interaction, the phonon
satellites are quite sensitive to the lead Fermi levels with
respect to the localized level.\cite{CoulumbFree} When the
temperature is too low to excite phonons thermally, the phonon
satellites can only develop below (above) the resonant level if
this level is occupied (empty) initially, as then electrons can
only emit phonons. On the other hand, in the Kondo regime with the
EPI ignored, it is well known that two collective spin singlet
states will be formed between the localized state and the excited
electron states above the Fermi level $\mu$ or the excited hole
states below $\mu$ because of spin exchange
processes,\cite{twosinglets1,twosinglets2} which will contribute
to the formation of the sharp Kondo peak pinning at the Fermi
surface in the local density of states. It is thus expected that
the interesting transport physics may result from the interplay of
both the EPI and the Kondo effect, in particular out of the
equilibrium, as both effects are sensitive to the Fermi levels at
the source and drain electrode.

Several theoretical techniques have been developed and applied to
address the joint effects of the EPI and the Kondo correlation on
transport through a quantum dot or a SMT.
\cite{SlaveBoson,NRGHewson,NRG1,NRG2,NRG3,SWtransformation,real-time1,real-time2,others1,others2,others3}
According to these works, the equilibrium properties, in
particular the renormalized effect on the Kondo correlation due to
the EPI, have been well described. For example, with the help of
the numerical renormalization group approach, the renormalized
effect has been predicted for wide range of parameters in the
equilibrium.\cite{NRGHewson,NRG1,NRG2,NRG3} With the help of
generalized Schrieffer-Wolff transformation, under the assumption
that the system stays equilibrium in the strongly asymmetric
coupling case, the nonlinear differential conductance has been
given in Ref.\cite{SWtransformation} with the Kondo satellite
structure exhibited. On the other hand, owing to the lack of
satisfactory treatment for the nonequilibrium Kondo problem,
little effort has been made to investigate how the Kondo
phonon-satellites develop and manifest themselves in the
nonequilibrium transport. However, since the Kondo
phonon-satellites can only be observed at finite bias voltage
experimentally, when the system is driven out of
equilibrium,\cite{KondoSattelites1,KondoSattelites2,KondoSattelites3}
it is certainly of importance to investigate the key signature of
the nonequilibrium phonon-Kondo effect. As far as we know, by the
real-time diagrammatic formulation, K$\ddot{o}$ning et al. have
given the spectral function as well as the differential
conductance with Kondo satellites in the nonequilibrium
situation.\cite{real-time1,real-time2} Here we shall develop
another easier and straightforward approach, which will combine
the improved nonperturbative canonical transformation treatment of
the EPI \cite{CoulumbFree} with the extended equation-of-motion
(EOM) method of the nonequilibrium Green functions. This approach,
although is not rigorous in quantitative description, has the
advantage of intuitiveness and can provide a semi-quantitative
understanding of the phonon-assisted Kondo effect, particularly
for strong electron-vibration coupling as well as nonequilibrium
situation.

Based on the approach mentioned above, in the present paper, we
shall mainly focus on how the Kondo satellites develop and
manifest themselves in the nonequilibrium transport through the
SMT, and how these Kondo satellites are affected by an applied
local magnetic field. After the Introduction we shall give our
model and method in Section II and III, respectively. Our main
results are presented in Section IV, which include: (i) The Kondo
phonon-satellites may exhibit quite different asymmetric
line-shapes with respect to the Kondo main peak in the spectral
function, and also in the nonlinear differential conductance; (ii)
Two types of spin exchange processes, associated with the
collective spin singlets formed by electron states and by hole
states respectively, can be clearly distinguished by the Kondo
satellites; and (iii) With a moderate Zeeman splitting exceeded
the width of Kondo peak, its Kondo-phonon satellites only appear
on one side of the main Kondo peak in one spin-resolved spectral
function, and on the opposite side for the opposite spin
component. Finally, the conclusions are drawn.

\section{Model}
The SMT system studied in the present paper can usually be
described as the Anderson-Holstein model, in which a single
localized state coupled linearly to one local vibration mode and
to the left (L) and right (R) non-interacting leads, namely
\begin{eqnarray}
{\bf H}&=&\sum_{\alpha;{\bf k},\sigma} \varepsilon_{\alpha\bf k}
{\bf c}^{\dag}_{\alpha{\bf k}\sigma}{\bf c}_{\alpha{\bf
k}\sigma}+\sum_{\sigma}\varepsilon_{\sigma}{\bf n}_{\sigma}
+U_0{\bf n}_{\uparrow}{\bf n}_{\downarrow} +\hbar\omega_0{\bf
a}^{\dag}{\bf a}\nonumber\\
&&+\sum_{\sigma}\lambda{\bf n}_{\sigma}({\bf a}^{\dag} + {\bf a})
+\sum_{\alpha;{\bf k},\sigma} (V_{\alpha{\bf k}} {\bf
c}^{\dag}_{\alpha{\bf k}\sigma} {\bf d}_{\sigma} + h.c.
),\label{H-previous}
\end{eqnarray}
where ${\bf c}^{\dag}_{\alpha{\bf k}\sigma}$(${\bf c}_{\alpha{\bf
k}\sigma}$) and ${\bf d}^{\dag}_{\sigma}$(${\bf d}_{\sigma}$) are
the creation (annihilation) operators for the lead electron with
energy $\varepsilon _{\alpha{\bf k}}$ and the localized SMT
electron with energy $\varepsilon_{\sigma}$, respectively,
$\sigma$ denotes the spin index and the lead index $\alpha=L,R$.
$U_0$ is the on-site Coulomb repulsion, and ${\bf n}_{\sigma}={\bf
d}^{\dag}_{\sigma}{\bf d}_{\sigma}$. The operator ${\bf a}^{\dag
}$ (${\bf a}$) creates (annihilates) the local vibration mode with
frequency $\omega_0$, $\lambda$ is the EPI strength,
$V_{\alpha{\bf k}}$ is the tunneling coupling between the
localized and lead electrons, which results in a level broadening
$\Gamma=(\Gamma_{L}+\Gamma_{R})/2$, where
\begin{eqnarray}
\Gamma_{\alpha}(\omega)\equiv 2\pi\sum_{\bf k}|V_{\alpha{\bf
k}}|^2\delta(\omega-\varepsilon_{\alpha{\bf k}}).
\end{eqnarray}

To treat the EPI non-perturbatively, by the canonical
transformation, $\mathcal{\bf S}=\frac{\lambda }{\omega
_{0}}\sum_{\sigma}{\bf n}_{\sigma}\left( {\bf a}^{\dag }-{\bf
a}\right)$, the Hamiltonian is transformed into $\bar{\bf H}\equiv
e^{\mathcal{\bf S}}{\bf H}e^{-\mathcal{\bf S}}=\bar{\bf
H}_{ph}+\bar{\bf H}_{el}$, where the phonon part $\bar{\bf
H}_{ph}=\hbar\omega_0{\bf a}^{\dag}{\bf a}$, and the electron part
turns out to be the Anderson Hamiltonian, namely
\begin{eqnarray}
\bar{\bf H}_{el}=\sum_{\alpha,{\bf k},\sigma}
\varepsilon_{\alpha\bf k} {\bf c}^{\dag}_{\alpha{\bf k}\sigma}{\bf
c}_{\alpha{\bf k}\sigma}+\sum_{\sigma}\bar{\varepsilon
}_{\sigma}{\bf
n}_{\sigma}+\bar{U}_0{\bf n}_{\uparrow}{\bf n}_{\downarrow}\nonumber\\
+\sum_{\alpha,{\bf k},\sigma} (\bar{V}_{\alpha{\bf k}} {\bf
c}^{\dag}_{\alpha{\bf k}\sigma} {\bf d}_{\sigma} + h.c.
).\label{H-el_bar}
\end{eqnarray}
The parameters with bar in Eq.(\ref{H-el_bar}) correspond to the
renormalized ones resulted from the canonical transformation, e.g.,
the SMT level is shifted to
$\bar{\varepsilon}_{\sigma}=\varepsilon_{\sigma}-g\omega_0$ with $
g\equiv\left( \lambda /\omega _{0}\right) ^{2}$, and the on-site
repulsion is renormalized to $\bar{U}_0=U_0-2g\omega_0$. When the
charging energy is significantly reduced by screening due to the
electrodes, \cite{screeningSMT} $\bar{U}_0$ may become negative for
very strong EPI, which will result in the anisotropic Kondo
effect.\cite{NRGHewson,NRG1,NRG2,NRG3} But in most realistic
situations, the charging energy is much larger than the EPI
energy,\cite{PhononSMT1,PhononSMT2,PhononSMT3} so we shall limit
ourselves to the standard Kondo effect, i.e. $\bar{U}_{0}$ remains
positive. Besides, $\bar{V}_{\sigma{\bf k}}\equiv V_{\sigma{\bf
k}}{\bf X}$ with ${\bf X}\equiv\exp \left[ -\left( \lambda /\omega
_{0}\right) \left( {\bf a}^{\dag }-{\bf a}\right) \right]$,
indicating that a cloud of phonons will be created or destroyed
during the hopping processes. As containing the phonon operator
$\bf{X}$, the renormalized Hamiltonian $\bar{\bf H}_{el}$ has not
been fully decoupled. For the present work, we are specially
interested in the cases where the EPI is sufficiently strong
compared to the tunneling coupling so that a local polaron is
formed. Then we can approximate the phonon operator ${\bf X}$ by its
expectation value in the thermal equilibrium, $\langle {\bf X}
\rangle=\exp \left[ -g\left( N_{ph}+1/2\right) \right]$, with
$N_{ph}$ defined as the population of phonons at temperature $T$.
Although it belongs to the mean field approximation, this method has
been used widely and obtained reasonable
successes.\cite{PreviousDecouple1,PreviousDecouple2,PreviousDecouple3,EPIDecoupling,polaron1,polaron2}

With this mean field approximations, the lesser Green function can
then be separated into the electron part and phonon part,
\begin{eqnarray}
&&G^{<}_{\sigma}\left( t\right) \equiv i\left\langle {\bf
d}_{\sigma}^{\dag }\left( 0\right) {\bf d}_{\sigma}\left( t\right)
\right\rangle=i\left\langle \overline{\bf d}_{\sigma}^{\dag
}e^{i\overline{\bf H}t} \overline{\bf
d}_{\sigma}e^{-i\overline{\bf H}t}
\right\rangle\nonumber\\
&=& i\left\langle {\bf d}_{\sigma}^{\dag }e^{i\overline{\bf
H}_{el}t} {\bf d}_{\sigma}e^{-i\overline{\bf H}_{el}t}
\right\rangle_{el}\left\langle {\bf X}^{\dag}e^{i\overline{\bf
H}_{ph}t}{\bf X}e^{-i\overline{\bf H}_{ph}t} \right\rangle_{ph}.
\end{eqnarray}
The trace of the phonon part can be evaluated by the Feynman
disentagling technique.\cite{EPIDecoupling} The lesser Green
function of the SMT electron can be decoupled as
\begin{eqnarray}
G^<_{\sigma}(\omega)
&=&\sum_{n=-\infty}^{\infty}L_n\bar{G}^<_{\sigma}(\omega+n\omega_0),\label{approximation_G^>}
\end{eqnarray}
and similarly the greater Green function is formulated as
\begin{eqnarray}
G^>_{\sigma}(\omega)=\sum_{n=-\infty}^{\infty}L_n\bar{G}^>_{\sigma}(\omega-n\omega_0).\label{G-greater}
\end{eqnarray}
Here $\bar{G}^{>(<)}_{\sigma}$ is the dressed Green function
associated with $\bar{\bf H}_{el}$, and
\begin{equation}
L_n= e^{-g\left( 2N_{ph}+1\right)
}e^{n\omega_0/2k_{B}T}I_{n}\left( 2g\sqrt{N_{ph}\left(
N_{ph}+1\right) }\right),
\end{equation}
in which $I_n(z)$ is the $n$-th Bessel function of complex
argument. The spectral function is then calculated with the lesser
and greater Green functions via
\begin{eqnarray}
A_{\sigma}(\omega)=i(G^{>}_{\sigma}(\omega)-G^{<}_{\sigma}(\omega)).
\end{eqnarray}
It should be emphasized that people usually evaluate the spectral
function via
$A_{\sigma}(\omega)=i(G^{r}_{\sigma}(\omega)-G^{a}_{\sigma}(\omega))$,
in which the retarded and advanced Green functions,
$G^{r}_{\sigma}$ and $G^{a}_{\sigma}$, are approximately separated
into the electron and phonon parts by ignoring the difference
between the $N_{ph}$ and $N_{ph}+1$, which works only in the
high-temperature
limit.\cite{PreviousDecouple1,PreviousDecouple2,PreviousDecouple3}
However, the phonon-Kondo problem studied here is certainly not
the case, as the characteristic Kondo temperature, $T_{K}$, is
very low. Hence the prescription we have recently made may be more
appropriate\cite{CoulumbFree}.

\section{Keldysh Green functions}

Within the Keldysh formalism, the dressed Green functions
$\bar{G}^{>(<)}_{\sigma}$ can be derived from $\bar{\bf H}_{el}$
with the help of the EOM approach. Since the usual approximation
in decoupling the higher-order Green functions generated by EOM is
valid only nearby or above the Kondo temperature
$T_{K}$,\cite{nonequilibrium1,nonequilibrium2} at lower
temperature as in the present situation the approximation
developed by Lacroix may be more
appropriate\cite{ImprovedDecoupling}. In the original Lacroix
approach, only the system with the spin degeneracy and infinite
charging energy was discussed in the equilibrium. To apply to the
nonequilibrium situation with the finite $U_0$ and Zeeman
splitting, we have extended it accordingly. Thus, the usually
ignored higher-order retarded Green function are taken into
reconsideration. For example, $\langle\langle{\bf
d}_{\bar{\sigma}}{\bf c}_{\alpha{\bf k}\bar{\sigma}}^{\dag}{\bf
c}_{\alpha'{\bf k}'\sigma},{\bf d}^{\dag}_{\sigma}\rangle\rangle$
is now approximately decoupled as $\langle{\bf
d}_{\bar{\sigma}}{\bf c}_{\alpha{\bf
k}\bar{\sigma}}^{\dag}\rangle\langle\langle{\bf c}_{\alpha'{\bf
k}'\sigma},{\bf d}^{\dag}_{\sigma}\rangle\rangle$, and the matrix
element $\langle{\bf d}_{\bar{\sigma}}{\bf c}_{\alpha{\bf
k}\bar{\sigma}}^{\dag}\rangle$ should be determined
self-consistently by the nonequilibrium Green function technique.

From the standard derivation of the EOM,\cite{footnote1,footnote2}
the dressed retarded Green function turns out to be
\begin{eqnarray}
\bar{G}^r_{\sigma}(\omega)=\frac{1-\left(\langle\bar{\bf
n}_{\bar{\sigma}}\rangle+\bar{P}_{\bar{\sigma}}(\bar{\omega}^0_{\bar{\sigma}})
+\bar{P}^*_{\bar{\sigma}}(-\bar{\omega}^{U})\right)}
{\omega-\bar{\varepsilon}_{\sigma}-\bar{\Sigma}^{(T)}(\omega)+
\frac{\bar{U}_0\bar{\Sigma}^{(1)}_{\bar{\sigma}}(\omega)}
{\omega-\bar{\varepsilon}_{\sigma}-\bar{U}_0-\bar{\Sigma}^{(T)}(\omega)
-\bar{\Sigma}^{(3)}_{\bar{\sigma}}(\omega)}}\nonumber\\
+\frac{\langle\bar{\bf
n}_{\bar{\sigma}}\rangle+\bar{P}_{\bar{\sigma}}(\bar{\omega}^0_{\bar{\sigma}})
+\bar{P}^*_{\bar{\sigma}}(-\bar{\omega}^{U})}
{\omega-\bar{\varepsilon}_{\sigma}-\bar{U}_0-\bar{\Sigma}^{(T)}(\omega)-
\frac{\bar{U}_0\bar{\Sigma}^{(2)}_{\bar{\sigma}}(\omega)}
{\omega-\bar{\varepsilon}_{\sigma}-\bar{\Sigma}^{(T)}(\omega)
-\bar{\Sigma}^{(3)}_{\bar{\sigma}}(\omega)}},\label{lesserG}
\end{eqnarray}
where
\begin{eqnarray}
\bar{\omega}^0_{\bar{\sigma}}&\equiv&\omega+\bar{\varepsilon}_{\bar{\sigma}}
-\bar{\varepsilon}_{\sigma},\nonumber\\
\bar{\omega}^{U}&\equiv&\omega-\bar{\varepsilon}_{\sigma}
-\bar{\varepsilon}_{\bar{\sigma}}-\bar{U}_0,\label{w0}
\end{eqnarray}
and the correlation function between the SMT level and leads,
$\bar{P}_{\bar{\sigma}}(\omega)$, is defined as
\begin{eqnarray}
\bar{P}_{\bar{\sigma}}(\omega)\equiv\sum_{\alpha{\bf
k}}\bar{V}^*_{\alpha{\bf k}}\langle{\bf
d}^{\dag}_{\bar{\sigma}}{\bf c}_{\alpha{\bf k}\bar{\sigma}}\rangle
g^r_{\alpha{\bf k}}(\omega),\label{p}
\end{eqnarray}
in which $g^r_{\alpha{\bf k}}(\omega)$ is the unperturbed retarded
Green function of the lead electron. In the above the self-energy
associated with the resonant tunneling processes is defined as
\begin{eqnarray}
\bar{\Sigma}^{(T)}(\omega)\equiv\sum_{\alpha{\bf
k}}|\bar{V}_{\alpha{\bf k}}|^2 g^r_{\alpha{\bf k}}(\omega),
\end{eqnarray}
and the self-energies related to the Kondo correlation are
respectively expressed as
\begin{eqnarray}
\bar{\Sigma}^{(1)}_{\bar{\sigma}}(\omega)&=&\bar{\Sigma}^{(0)}_{\bar{\sigma}}(\omega)-\bar{\Sigma}^{(T)}(\omega)
\left[\bar{P}_{\bar{\sigma}}(\bar{\omega}^0_{\sigma})+\bar{P}^*_{\bar{\sigma}}(-\bar{\omega}^{U})\right],\nonumber\\
\bar{\Sigma}^{(3)}_{\bar{\sigma}}(\omega)&=&\bar{\Sigma}^{(T)}(\bar{\omega}^0_{\bar{\sigma}})
-\left(\bar{\Sigma}^{(T)}(-\bar{\omega}^{U})\right)^*,\nonumber\\
\bar{\Sigma}^{(2)}_{\bar{\sigma}}(\omega)&=&\bar{\Sigma}^{(3)}_{\bar{\sigma}}
(\omega)-\bar{\Sigma}^{(1)}_{\bar{\sigma}}(\omega).
\end{eqnarray}
Here
$\bar{\Sigma}^{(0)}_{\bar{\sigma}}(\omega)=\bar{F}_{\bar{\sigma}}(\bar{\omega}^0_{\bar{\sigma}})
-\bar{F}^*_{\bar{\sigma}}(-\bar{\omega}^{U})+\bar{B}_{\bar{\sigma}}(\omega)$,
\begin{eqnarray}
\bar{F}_{\bar{\sigma}}(\omega)=\sum_{\alpha{\bf k},\alpha'{\bf
k}'}\bar{V}^*_{\alpha{\bf k}}\bar{V}_{\alpha'{\bf k}'}\langle{\bf
c}^{\dag}_{\alpha'{\bf k}'\bar{\sigma}}{\bf c}_{\alpha{\bf
k}\bar{\sigma}}\rangle g^r_{\alpha{\bf k}}(\omega),\label{F}
\end{eqnarray}
and
\begin{eqnarray}
\bar{B}_{\bar{\sigma}}(\omega)=-2iIm\left\{\sum_{\alpha'{\bf
k}'}\bar{V}^*_{\alpha'{\bf k}'}\left\langle{\bf
d}^{\dag}_{\bar{\sigma}}{\bf c}_{\alpha'{\bf
k}'\bar{\sigma}}\right\rangle\right\}\nonumber\\
\times\left[\sum_{\alpha{\bf k}}|\bar{V}_{\alpha{\bf
k}}|^2\left(g^r_{\alpha{\bf k}}(\omega)\right)^2\right].
\end{eqnarray}
In the wide-band limit,
$\bar{\Sigma}^{(T)}(\omega)\approx-i\bar{\Gamma}$, and
\begin{eqnarray}
\bar{\Sigma}^{(0)}_{\bar{\sigma}}(\omega)\approx\sum_{\alpha}\frac{\bar{\Gamma}_{\alpha}}{2\pi}
\left[\psi\left(\frac{1}{2}+\frac{\bar{\omega}^{U}+\mu_{\alpha}}{i2\pi
k_{B}T}\right)\right.\nonumber\\
\left.-\psi\left(\frac{1}{2}+\frac{\bar{\omega}^0_{\bar{\sigma}}-\mu_{\alpha}}{i2\pi
k_{B}T}\right)-i\pi\right],
\end{eqnarray}
where the Psi function $\psi$ is the logarithmic derivative of the
gamma function. Now the self-consistent equation for
$\bar{P}_{\bar{\sigma}}(\omega)$ is simplified as
\begin{eqnarray}
\bar{P}_{\bar{\sigma}}(\omega)=\int \frac{d\omega'}{2\pi}
\frac{[\bar{\Gamma}_Lf_L(\omega')+\bar{\Gamma}_Rf_R(\omega')]\bar{G}^{r*}_{\bar{\sigma}}(\omega')}{\omega-\omega'+i0^+},
\end{eqnarray}
which, together with Eq.(\ref{lesserG}), gives the dressed
retarded Green function $\bar{G}^{r}_{\sigma}$.

As we will see that at zero temperature, not only
$\bar{F}_{\bar{\sigma}}(\omega)$, but also
$\bar{P}_{\bar{\sigma}}(\omega)$ will diverge logarithmically and
then all contribute to the sharp Kondo resonance at each Fermi
surface. Therefore, $\bar{P}_{\bar{\sigma}}(\omega)$ should not be
ignored as in the previous
approaches.\cite{nonequilibrium1,nonequilibrium2} Generally, this
EOM approach works for most cases, however it fails to give the
Kondo resonance at the particle-hole symmetric case. For example,
when
$\bar{\varepsilon}_{\uparrow}=\bar{\varepsilon}_{\downarrow}=-\bar{U}_0/2$,
Eqs.(\ref{w0},\ref{p},\ref{F}) show that
$\bar{\omega}^0_{\bar{\sigma}}=\bar{\omega}^U$,
$\bar{P}_{\bar{\sigma}}^*(-\omega)=-\bar{P}_{\bar{\sigma}}(\omega)$,
and
$\bar{F}_{\bar{\sigma}}^*(-\omega)=-\bar{F}_{\bar{\sigma}}(\omega)$.
Therefore the divergent terms in the expression of the retarded
Green function are cancelled with each other and then no Kondo
resonance will show up.\cite{footnote3} The reason for this
limitation is the fact that at the particle-hole symmetry point,
the charge fluctuations are completely quenched, therefore the EOM
approach can not be applied.\cite{FerromaneticLeads2}

By the Keldysh formula
$\bar{G}^{>(<)}_{\sigma}=\bar{G}^r_{\sigma}\bar{\Sigma}^{>(<)}_{\sigma}\bar{G}^a_{\sigma}$,
in which the the greater (lesser) self-energy is evaluated by the
ansatz adopted by Ng\cite{NgAnzatz}, the dressed greater and
lesser Green functions are obtained as
\begin{eqnarray}
\bar{G}^>_{\sigma}(\omega)&=&i([\bar{\Gamma}_L(1-f_L(\omega))
+\bar{\Gamma}_R(1-f_R(\omega))])Im\bar{G}^r_{\sigma}(\omega)/\bar{\Gamma},\nonumber\\
\bar{G}^<_{\sigma}(\omega)&=&-i[\bar{\Gamma}_Lf_L(\omega)
+\bar{\Gamma}_Rf_R(\omega)]Im\bar{G}^r_{\sigma}(\omega)/\bar{\Gamma}.\label{renormalizedGgreater}
\end{eqnarray}
Substituting Eq.(\ref{renormalizedGgreater}) into
Eqs.(\ref{approximation_G^>}) and (\ref{G-greater}), one obtains
the greater (lesser) Green function and subsequently the spectral
function of the SMT.

It is noted that, with the infinity charging energy and zero
Zeeman splitting in the equilibrium, our results will recover to
the Lacroix's one.\cite{ImprovedDecoupling} Together with the
improved canonical transformation, this generalized EOM method
will be our framework for investigating the Kondo satellites at
the nonequilibrium with finite Zeeman splitting. Although this
approach can only give a qualitative description of the shape of
the Kondo satellites, it does predict the right peak
positions.\cite{nonequilibrium1,nonequilibrium2,NRGHewson} For
quantitative description of the renormalized properties in the
equilibrium or exploring wider parameter space, more accurate
methods like the NRG or quantum monte carlo calculations may be
appropriate\cite{NRGHewson,NRG1,NRG2,NRG3}.

With the spectral function and the lesser Green function given
above, the current through the SMT can be calculated
via\cite{Transport1,Transport2}
\begin{eqnarray}
J=\sum_{\sigma}\frac{ie}{2h}\int d\omega \{[ f_{L}( \omega
)\Gamma_{L}-f_{R}( \omega )\Gamma _{R}]A_{\sigma}(\omega)
\nonumber\\
+( \Gamma_{L}-\Gamma_{R})G^<_{\sigma}(\omega) \},
\end{eqnarray}
and the differential conductance can be obtained straightforward
as $G=\partial J/\partial V_{bias}$.

\section{Spectral Function and Differential Conductance}

\subsection{spin degenerate case}

\begin{figure}[tbp]
\begin{centering}
\includegraphics[width=8.5cm]{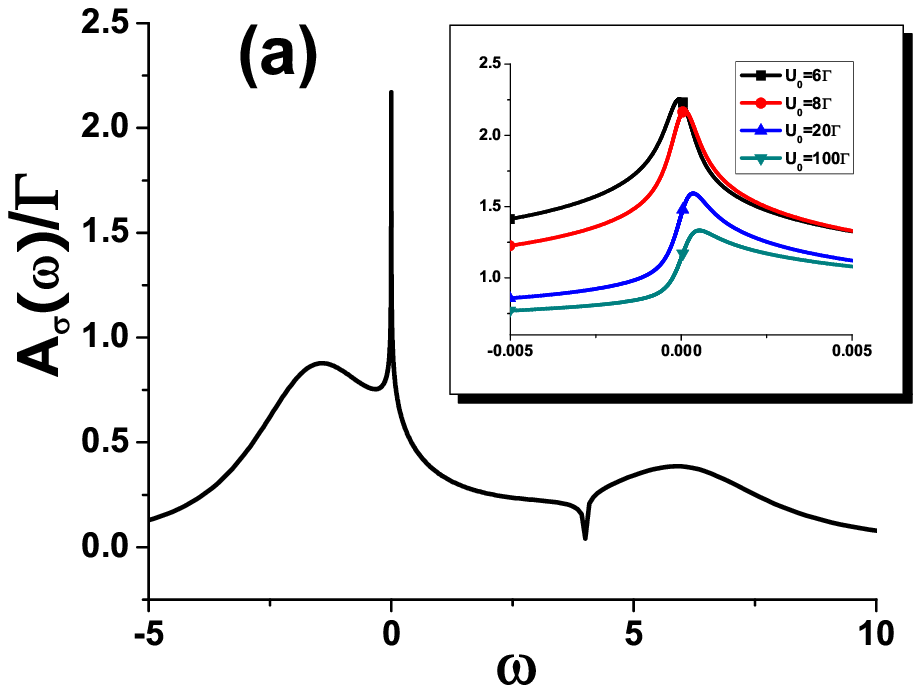}
\includegraphics[width=8.5cm]{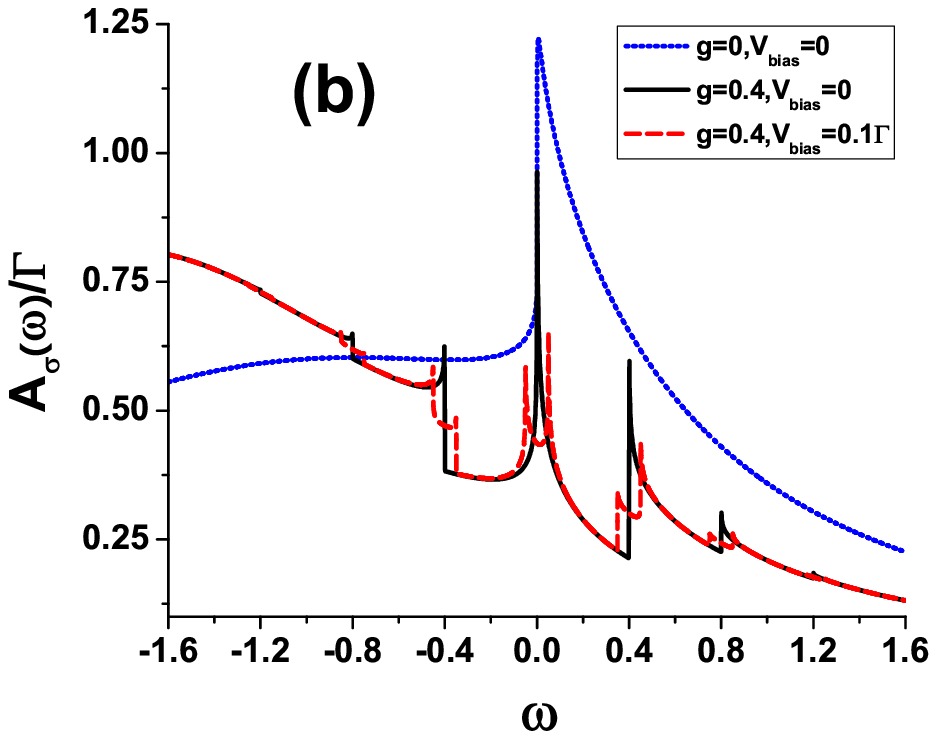}
\caption{(color online) The spectral functions of the spin
degenerate local electrons, i.e., $A_{\sigma}$, with (a) and
without (b) the EPI. The parameters used in (a) are:
$\mu_L=\mu_R=0.0$, $\varepsilon_0=-2\Gamma$, and $U_0=8\Gamma$,
$V_{bias}=0$. The parameters used in (b) are
$(\mu_{L}+\mu_{R})/2=0$, $\varepsilon_0=-2.5\Gamma$,
$k_{B}T=2\times10^{-4}\Gamma$, $U_0=100\Gamma$,
$\hbar\omega_0=0.4\Gamma$. Other parameters have been shown
explicitly in each figures.} \label{spectral}
\end{centering}
\end{figure}

First, we will discuss the spin degenerate case, i.e.,
$\varepsilon_{\uparrow}=\varepsilon_{\downarrow}=\varepsilon_0$.
The spectral function in the absence of the EPI is plotted in
Fig.\ref{spectral}(a) for different $\bar{U}_0$, which shows that
besides the two broad resonant peaks at $\bar{\varepsilon}_0$ and
$\bar{\varepsilon}_0+\bar{U}_0$, there will be a sharp Kondo
resonance developed at $\mu_{L(R)}$. By increasing $\bar{U}_0$, as
can be observed from the inset of Fig.\ref{spectral}(a), this
Kondo resonant peak can be tuned more and more asymmetric, which
is a manifestation of the particle-hole asymmetry of the SMT.

When the electron-phonon interaction is turned on, as shown in
Fig.\ref{spectral}(b), the Kondo peak will be split into a set of
phonon satellites spaced by multiple of phonon frequency. When the
system is driven out of equilibrium by applied a bias voltage, the
main Kondo peak will be split into two weakened peaks pinning at
Fermi levels of the left and right lead,respectively, and the
phonon Kondo satellites will also be split into two corresponding
sub-sets. These results well agree with previous works, which
validates the present
framework\cite{nonequilibrium1,nonequilibrium2,real-time1,real-time2}.

\begin{figure}[tbp]
\begin{centering}
\includegraphics[width=8.5cm]{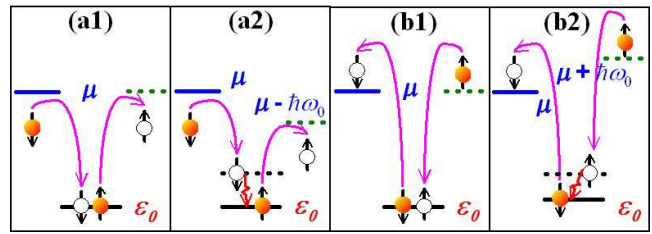}
\caption{(color online) Schematic illustration for the spin
exchange processes associated with the formation of the Kondo
satellites.} \label{SpinExchange}
\end{centering}
\end{figure}

The underlying physics about how and why the Kondo satellites
develop in such a way can be intuitively understood with the help of
Fig.\ref{SpinExchange}. As illustrated by
Fig.\ref{SpinExchange}(a1), a local hole state in the SMT exchanges
with the excited holes at leads below the Fermi energy $\mu$ to form
a Kondo spin-singlet. When an electron hops to the local state by
emitting a phonon, as shown in Fig.\ref{SpinExchange}(a2), only the
holes with energy below $\mu-\hbar\omega_0$ can exchange with the
local hole, because the phonon absorption is unavailable at very low
temperature and the energy conservation law must be satisfied in the
interaction process. Thus a phonon-satellite peak at
$\mu-\hbar\omega_0$ appears in the density of states or tunneling
spectra. Similarly, a spin-singlet consisting of a localized
electron and excited electrons above $\mu$ is depicted in
Fig.\ref{SpinExchange}(b1), and only the electrons above
$\mu+\hbar\omega_0$ can exchange with the local electron if the
hopping processes are accompanied by emitting a phonon as shown in
Fig.\ref{SpinExchange}(b2), which will give rise to the satellite
peak at $\mu+\hbar\omega_0$. That means although both processes
shown in Fig.\ref{SpinExchange}(a1) and Fig.\ref{SpinExchange}(b1)
contribute to the main Kondo peak, they can be distinguished by
their Kondo satellites, namely, the contributions of each type of
exchange processes to the Kondo peak can be picked out in principle
by comparing with the satellites on each sides of the main peak.

Fig.\ref{spectral}(b) also shows the satellites on different sides
of the Kondo main peak may take on apparent sharp asymmetric
lineshapes. The reason for these sharp feature lies in the sharp
Fermi distributions in the leads at very low temperature, where
the electron excitations and the hole excitations are separated
from each other sharply by the Fermi surface. For example, the
phonon-assisted spin exchange processes mentioned above have the
sharp energy boundaries as $\mu_{L(R)}\pm n\omega_0$, therefore
they may result in the sharp features at $\omega=\mu_{L(R)}\pm
n\omega_0$ as shown in Fig.\ref{spectral}(b). Besides, if the
localized level is broadened so that there is a finite density of
states around $\mu_{L(R)}$, the electron can also bypass the SMT
via the phonon-assisted resonant tunneling processes. According to
our previous studies,\cite{CoulumbFree} these phonon-assisted
resonant tunneling processes can result in the phonon sidebands at
$\omega=\bar{\varepsilon}_0\pm n\omega_0$, which also exhibit some
sharp asymmetric lineshapes at $\omega=\mu_{L(R)}\pm n\omega_0$
under extremely low temperature. These structures superpose over
the Kondo-phonon-satellites, jointly responsible for the apparent
sharp asymmetric peaks observed in Fig.\ref{spectral}(b).

\begin{figure}[tbp]
\begin{centering}
\includegraphics[width=8.5cm]{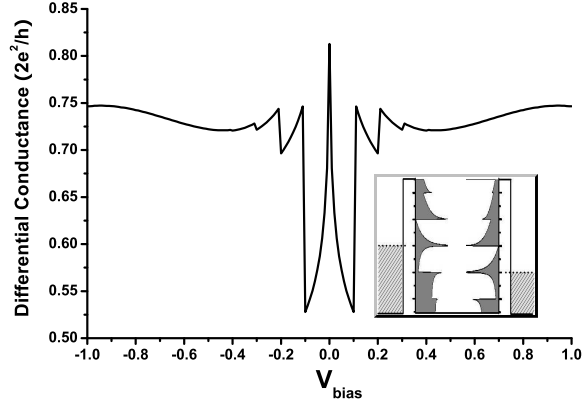}
\caption{The differential conductance for spin degenerate case as
a function of the bias voltage. Here
$\bar{\epsilon}_0=-2.5\Gamma$, $g=0.5$,
$k_{B}T=2\times10^{-4}\Gamma$, $U_0=100\Gamma$, and
$\hbar\omega_0=0.1\Gamma$. As an illustration, the inset shows the
origin of the first Kondo phonon-satellite peak schematically.}
\label{DC}
\end{centering}
\end{figure}

The nonlinear differential conductance versus the bias voltage,
i.e., $G(V_{bias})$, is plotted in Fig.\ref{DC} for very low
temperature, where $G$ reaches a peak value at $V_{bias}=0$ and
then decreases rapidly as $V_{bias}$ increases. This zero-bias
anomaly is usually regarded as the experimental evidence for the
Kondo effect in the quantum dot systems. When
$eV_{bias}=n\hbar\omega_{0}$, there are some
Kondo-phonon-satellite peaks show up, as evidenced by the very
recent
experiments.\cite{KondoSattelites1,KondoSattelites2,KondoSattelites3}
According to the general understanding, in the non-EPI case the
bias voltage plays a role similar to temperature, which means the
differential conductance decreases logarithmically as $V_{bias}$
increases,\cite{Appelbaum66} one may wonder here why the apparent
sharp asymmetric features of the Phonon-Kondo-peaks in
Fig.\ref{DC} have not been smoothed out by bias voltage. It is
because the Fermi distributions in the leads have sharp Fermi
surfaces at very low temperature, the sharp features of the
spectral function, as shown in Fig.\ref{spectral}(b), can still be
preserved even for finite bias voltage. As illustrated in the
inset of Fig.\ref{DC}, once an asymmetric peak in the spectral
function matches one of the Fermi levels, a sharp asymmetric
differential conductance peak will appear. Therefore, the apparent
sharp asymmetric features in $G(V_{bias})$ are just the
manifestations of the asymmetric lineshapes of the spectral
function, which contain both contributions from the
phonon-assisted Kondo processes and the phonon-assisted resonant
tunneling processes as well. These sharp asymmetric features can
only be smoothed out with the increasing temperature.

It is worth pointing out the resemblances between our spin
degenerate results and those by the real-time diagrammatic
formulation\cite{real-time1,real-time2} or the generalized
Schrieffer-Wolff transformation\cite{SWtransformation}, where the
discontiguous and asymmetric lineshape features of the
Kondo-phonon-satellites in the spectral function or the
differential conductance can also be observed, but have not yet
been clearly pointed out and properly explained previously.

\subsection{finite Zeeman splitting case}
Now let us take a look at the spin non-degenerate case, where a
Zeeman splitting of $\Delta$ for the localized level may be induced
by a local magnetic field, i.e.
$\varepsilon_{\uparrow}=\varepsilon_{0}+\Delta/2$ and
$\varepsilon_{\downarrow}=\varepsilon_{0}-\Delta/2$.

\begin{figure}[tbp]
\begin{centering}
\includegraphics[width=8.5cm]{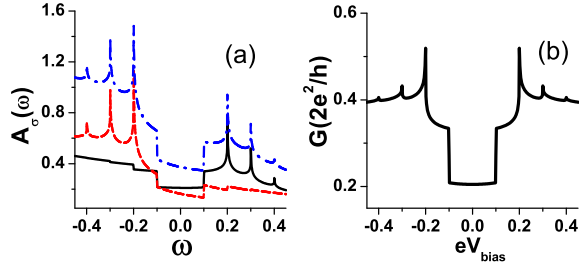}
\caption{(color online)  The calculated total (blue dash dot line)
and spin-resolved (black solid line for spin up, and red dash line
for spin down) spectral functions in the equilibrium (a), and
 the nonlinear differential conductance (b) in the presence of
finite Zeeman splitting. Here $\bar{\varepsilon}_{0}=-2.5\Gamma$,
$g=0.6$, $k_{B}T=2\times10^{-4}\Gamma$, $U_0=100\Gamma$,
$\hbar\omega_0=0.1 \Gamma$, and $\Delta=0.2\Gamma$.
\label{zeeman}}
\end{centering}
\end{figure}

Without the EPI, our calculation for the spin-splitting case shows
that, compared to the spin-degeneracy case, the Kondo peak in the
spin-resolved spectral function $A_{\uparrow}(\omega)$ (or
$A_{\downarrow}(\omega)$) generally decreases in magnitude and
shifts away from $\mu$ by $\Delta$ ($-\Delta$), which agrees with
previous
researches.\cite{real-time1,real-time2,NRGHewson,nonequilibrium1,nonequilibrium2,
NRG1,NRG2,NRG3}  When the EPI turns on, as shown in
Fig.\ref{zeeman}(a), the phonon satellite structure may develop in a
distinct way, that is, for moderate Zeeman splitting only the
satellites above (below) the main peak appear in
$A_{\uparrow}(\omega)$ ($A_{\downarrow}(\omega)$).

This peculiar feature in the spin-resolved spectral function stems
from the interplay between the EPI and Kondo effect in the
presence of the finite spin-splitting. It also manifests itself in
the differential conductance. As shown in Fig.\ref{zeeman}(b),
there is no satellite appearing in the region
$\mu-\Delta<eV_{bias}<\mu+\Delta$, which is generally true as long
as the system is at very low temperature. According to our simple
picture above ({\it cf} Fig.\ref{SpinExchange}), it is easy to
understand.

\begin{figure}[tbp]
\begin{centering}
\includegraphics[width=8.5cm]{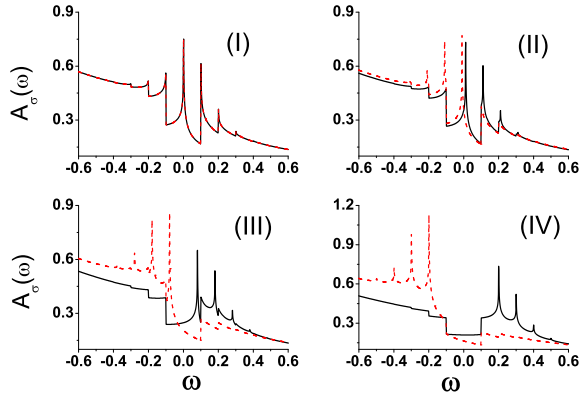}
\caption{(color online) The spin-resolved spectral function (black
solid line for spin up, red dash line for spin down component) for
different Zeeman splitting:(I) $\Delta=0$, (II)
$\Delta=0.01\Gamma$, (III) $\Delta=0.08\Gamma$, and (IV)
$\Delta=0.2\Gamma$. Other parameters are taken as $V_{bias}=0$,
$\bar{\varepsilon}_{0}=-2.5\Gamma$, $g=0.6$, $U_0=100\Gamma$, and
$\hbar\omega_0=0.1 \Gamma$. \label{zeemanChanging}}
\end{centering}
\end{figure}

The spin-resolved spectral functions for four different Zeeman
splitting values are plotted in Fig.\ref{zeemanChanging},
demonstrating that when the Zeeman splitting turns on, the overall
Kondo peaks in the spin down spectral function shift towards the low
energy, while those for spin up make a blue shift. The Kondo
satellites above (below) the main peak in the spin down (up)
spectral function decrease significantly as the Zeeman splitting is
increased, then disappear when the $\Delta$ exceeds the line-width
of the Kondo peak which is proportional to the Kondo temperature
$k_{B}T_{K}$. However, for larger splitting, the main Kondo peak
itself will be much suppressed, so that this spin-dependence of the
Kondo phonon satellites will not be resolved at all.

To understand how this pattern develops, it is helpful to recall
that the satellite on different sides of the main peak is
associated with different spin exchange process as depicted in
Fig.\ref{SpinExchange}. This peculiar pattern for finite Zeeman
splitting indicates that the dominant contribution to the main
peak in $A_{\uparrow}(\omega)$ comes from the coupling of the
local state with the excited conduction electron states, while for
$A_{\downarrow}(\omega)$ the dominant contribution comes from the
coupling with the excited hole states. If $\Delta=0$, as
illustrated in Fig.\ref{SpinExchange}, both spin exchange
processes contribute to the Kondo main peak. When $\Delta$ starts
to increase, both spin exchange processes decay in magnitude, but
at different rates. When Zeeman splitting increases further, it
would be possible that while one spin exchange process remains
finite, the other is already totally suppressed. This results in
the spin separation of the Kondo phonon satellites in the spectral
function. However, when $\Delta\gg k_BT_K$, so that both processes
are unlikely to happen and finally the Kondo peaks are quenched.

\begin{figure}[tbp]
\begin{centering}
\includegraphics[width=4.0cm]{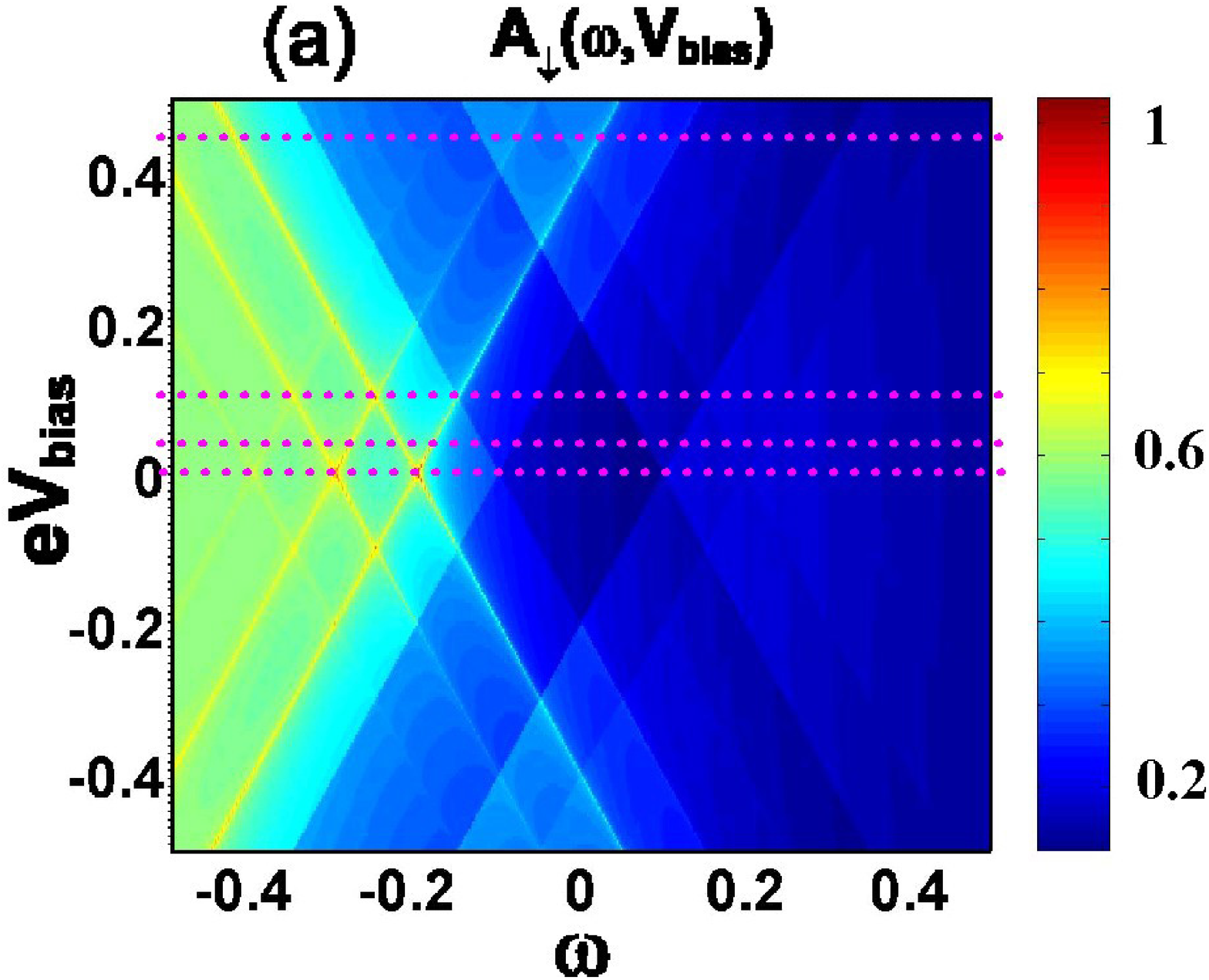}
\includegraphics[width=4.0cm]{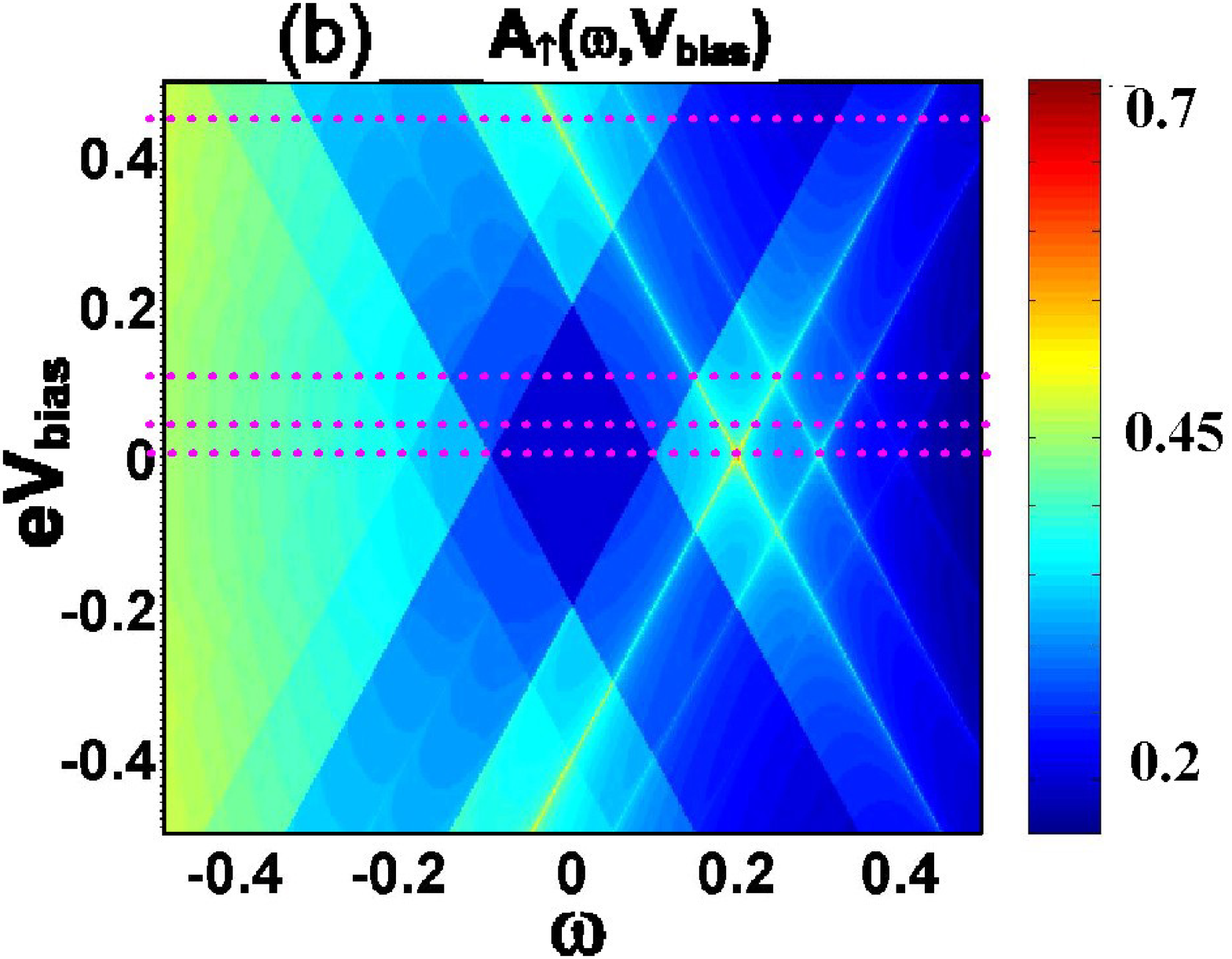}
\includegraphics[width=8.5cm]{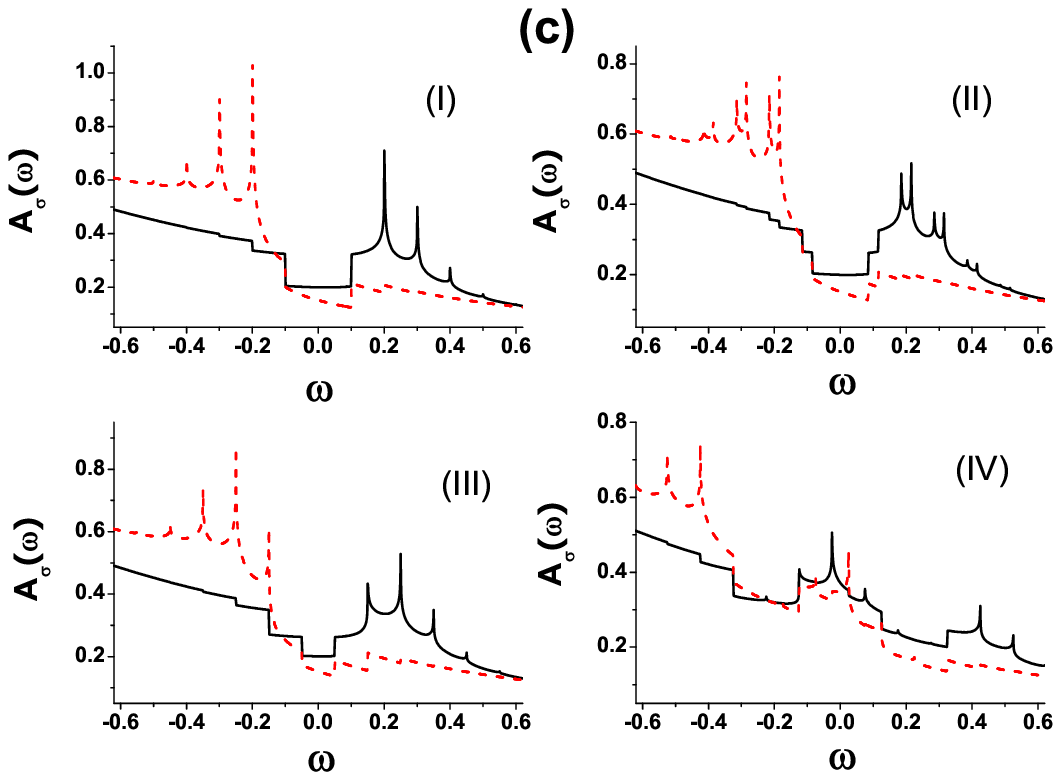}
\caption{(color online) Maps for the spin up (a) and spin down (b)
spectral functions as functions of $\omega$ and $eV_{bias}$. The
sections are plotted in (c) for four typical bias voltages: (I)
$eV_{bias}=0$, (II) $eV_{bias}=0.03\Gamma$, (III)
$eV_{bias}=0.1\Gamma$, (IV) $eV_{bias}=0.45\Gamma$. Here
$\Delta=0.2\Gamma$, $\bar{\varepsilon}_{0}=-2.5\Gamma$, $g=0.6$,
$U_0=100\Gamma$, and $\hbar\omega_0=0.1 \Gamma$.
\label{zeemanVbiasChanging}}
\end{centering}
\end{figure}

The spin down (up) spectral function for fixed Zeeman splitting is
shown in Fig.\ref{zeemanVbiasChanging}(a)
(Fig.\ref{zeemanVbiasChanging}(b)) as function of $\omega$ and
$V_{bias}$, in which the highlighting cross lines correspond to
the Kondo satellites. For explicitly, the sections at the four
typical bias voltages marked by the dash lines in
Fig.\ref{zeemanVbiasChanging}(a) and
Fig.\ref{zeemanVbiasChanging}(b) are also plotted in
Fig.\ref{zeemanVbiasChanging}(c), respectively. One can clearly
see that, when the bias voltage turns on, each Kondo peak is
separated into two sub-peaks related to two Fermi levels,
respectively. These two sub-sets of Kondo satellites separate from
each other further when $V_{bias}$ increases, resulting in the
peculiar pattern as shown in Fig.\ref{zeemanVbiasChanging}(c).

\section{Conclusion}

In summary, taking advantage of the improved treatment of the
electron-phonon interaction and the extended equation of motion
approach, we have systematically investigated the Kondo effect in
the SMT in the presence of the electron-phonon interaction, finite
bias voltage, and Zeeman splitting. The phonon-Kondo satellites
observed in the recent SMT transport
experiments\cite{KondoSattelites1,KondoSattelites2,KondoSattelites3}
have been confirmed and explained by our results. Moreover,
peculiar patterns of phonon-Kondo satellites in the spectral
function as well as the differential conductance have also been
predicted and explained by a clear physics picture. Although it is
the interplay between the EPI and Kondo processes that discussed
here, the physics picture and the theoretical calculation
presented here may also be applicable to some other cases. For
instance, the interplay of electronic internal excitations and the
Kondo processes can also manifest themselves in some parallel
Kondo satellites.\cite{Nygard00}

\ack The authors thank the helpful discussions with H. Zhai, Z. G.
Zhu, F. Ye, C. X. Liu and R. B. Liu. This work is supported by the
NSF of China (Grant No. 10374056, 10574076), the MOE of China
(Grant No. 200221), and the Program of Basic Research Development
of China (Grant No. 2001CB610508).


\begin{thebibliography}{10}
\bibitem{footnote3} The drawback of the EOM at the particle-hole
symmetic point has also been dicussed in Ref. \cite{footnote2}.
\end{thebibliography}
\end{document}